\documentclass[10pt,twocolumn,twoside]{IEEEtran}

\usepackage{graphicx,epsfig}
\usepackage{cite}
\usepackage{stfloats}
\usepackage{amsmath}
\usepackage{amsfonts,helvet}

\newtheorem{theorem}{Theorem}

\newtheorem{algo}[theorem]{Algorithm}

\newtheorem{proposition}[theorem]{Proposition}

\newcounter{proposition}
\begin{document}

\title{Novel Modulation Techniques using Isomers as Messenger Molecules for Molecular Communication via Diffusion}


\author{\authorblockN{Na-Rae Kim and Chan-Byoung Chae}\\
\authorblockA{School of Integrated Technology\\
Yonsei University, Korea\\
Email: \{nrkim, cbchae\}@yonsei.ac.kr\\
\thanks{This work was in part supported by the Ministry of Knowledge Economy under the ``IT Consilience Creative Program" (NIPA-2010-C1515-1001-0001) and the Yonsei University Research Fund of 2011.}}}




\maketitle \setcounter{page}{1} 
%
%
%


\begin{abstract}
In this paper, we propose novel modulation techniques using isomers as messenger molecules for nano communication via diffusion. To evaluate achievable rate performance, we compare the proposed techniques with concentration-based and molecular-type-based methods. Analytical and numerical results confirm that the proposed modulation techniques achieve higher data transmission rate performance than conventional insulin based concepts.   
\end{abstract}

\begin{keywords}
Nano communication network, molecular communication, modulation technique, isomer, diffusion, messenger molecule.
\end{keywords}

\section{Introduction}
Since R. Feynman's talk in 1959, people have eagerly pursued practical work on smaller and smaller scales, notably through nanotechnology~\cite{Feynman}. Nanotechnology, recently, produced a new branch of research called nano communication networks (NCNs). NCNs interconnect several nanoscale machines (nanomachines in short) to carry out more complex tasks in a cooperative manner, or to perform simple tasks~\cite{Akyildiz_cn08}. These networks are not just smaller versions of traditional communication networks; they have their own features and are applicable in many fields, including biomedical, industrial, military, and environmental~\cite{Akyildiz_cn09}.

NCNs can be realized by several methods.
We can rely on traditional communication systems that use electromagnetic fields or ultrasonic wave. We have to overcome, however, some radio frequency (RF) device barriers~\cite{Akyildiz_cn08}. On the practical front, researchers are considering new materials, such as carbon-nano tubes (CNTs) and graphene~\cite{Akan_commmag10}. Little research, however, has focused on channel models and human body absorption with terahertz bands. In addition to theses activities, a new concept utilizing diffusion especially for short range communication has been introduced~\cite{Moore_CPC06}. 

This new paradigm of communication, molecular communication, sends/receives information-encoded molecules between nanomachines. 
One of its advantages is its biocompatibility. Biocompatibility is the most important and challenging issue confronting intra-body applications. By using biomolecules we can create inherently biocompatible systems that require no inorganic harmful materials. Moreover, driven by chemical reactions, molecular communication is  energy efficient. For these reasons, molecular communication is the main focus of this paper.

The authors in~\cite{Kuran_NCN10,Kuran_ICC11} have extensively studied the fundamentals of molecular communication via diffusion. A new energy model has been investigated to understand how much energy is required to generate messenger molecules in~\cite{Kuran_NCN10} and concentration-based and molecular-type-based modulation techniques have been introduced in~\cite{Kuran_ICC11}. By using a simple symmetric binary channel model, the achievable transmission rate (achievable rate hereafter) has been extensively compared. In analyzing their modulation techniques, however,~\cite{Kuran_NCN10,Kuran_ICC11} did not clearly suggest concrete structures for the messenger molecules. They considered insulin-based nano networks but it is unclear how these could be utilized for molecular-type-based modulations. In this paper, to maximize the achievable rate with less transmit power/energy, we propose using isomers as messenger molecules. To consider the properties of isomers, we slightly modify the energy model described in~\cite{Kuran_NCN10}. To the best of our knowledge, the proposed method is the first attempt to design appropriate messenger molecules for NCN via diffusion. 

This paper is organized as follows. Section~II describes the channel and energy models under consideration. Section~III explains isomers for messenger molecules and also proposes two modulation techniques, i) concentration-based and ii) molecular-type-based. We present the numerical results in Section IV, and Section V offers our conclusions.

\section{System Model}
\label{Sec:Main} 
\begin{figure}[t]
 \centerline{\resizebox{0.9\columnwidth}{!}{\includegraphics{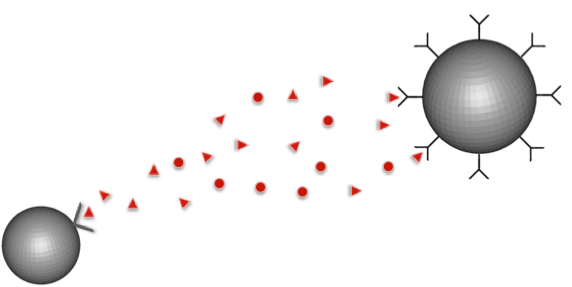}}}
  \caption{System model of nano communication via diffusion with a single transmitter and a single receiver.}
  \label{Fig:SysModel}
\end{figure}

For simplicity, we consider a nano communication system consisting of a single transmitter and a single receiver, as illustrated in Fig.~\ref{Fig:SysModel}. For this system, the messenger molecules diffuse through the medium\footnote{In our future work, blood will be considered as the medium to get more reasonable results.}~(i.e. water) at body temperature. This paper assumes that no collisions occur among the propagating messenger molecules and/or the molecules and the medium. 

\subsection{Channel Model}
The particles or messenger molecules released from a transmitter nanomachine spread out through the medium by Brownian motion~\cite{Kuran_NCN10}. Such motion is basically driven by diffusion, meaning the particles move from areas of higher concentration region to areas of lower concentration. A random process, the displacement of messenger molecules follows a normal distribution with zero mean. Thus, the standard deviation $\sigma$ of the displacement can be obtained as follows:
\begin{align}
&\bigtriangleup{X} \sim \mathcal{N}(0, \sigma^2), \nonumber \\
&\frac{\partial{c}}{\partial{t}}=D\frac{{\partial^2}c}{\partial{x^2}},\label{Fick's}\\
&c(x,t)=\frac{1}{(4{\pi}Dt)^{1/2}}e^{-x^2/{4Dt}},\label{c}\\
&\overline{x^2} = 2Dt,\nonumber\\
&\sigma=\sqrt{2Dt},\nonumber\\
& D = \frac{K_b T}{6\pi \eta r_{mm}} \label{D}
\end{align}
where, $c$ indicates the concentration of Brownian particles at time $t$ at point $x$, and $D$ represents the diffusion coefficient of the particles calculated from the Boltzmann constant ($K_b$), temperature ($T$), the viscosity of the medium ($\eta$) and the radius of a messenger molecule ($r_{mm}$). Eq. (\ref{Fick's}) is Fick's second law of diffusion. By solving this partial differential equation, we obtain the general equation for the concentration of the particles represented in (\ref{c}). The first moment of (\ref{c}) is zero, which indicates the displacement has a zero mean, and the second moment becomes the variance. Hence, the displacement has a standard deviation of $\sqrt{2Dt}$, and finally we have
\begin{align}
\bigtriangleup{X} \sim \mathcal{N}(0, 2Dt).
\end{align}

When $n$ messenger molecules are transmitted by the transmitter nanomachine, the molecules have a probability of hitting the receiver nanomachine. We represent this as a binomial distribution of $n$ times of trials with a probability of $P_{hit}$ for each. $P_{hit}$ is determined by the symbol duration ($T_s$) and the distance between the transmitter and the receiver ($d$), which are both affected by the diffusion coefficient. If $n$ is large enough, and $nP_{hit}$ is not zero, binomial can be approximated as a normal distribution. In addition, we have to take into account the overflow molecules from the previous symbol \cite{Kuran_NCN10}. The total number of molecules during a symbol duration becomes different depending on the current and previous symbol.
\begin{align}
&Binomial (n, P_{hit}(d,T_s)),\nonumber\\ 
&N_c  \sim \mathcal{N}(np_{1}, np_{1}(1-p_{1})),\nonumber\\
&N_p \sim \mathcal{N}np_2,np_2(1-p_2))-\mathcal{N}(np_1,np_1(1-p_1))
\nonumber  \end{align}
where, $N_c$ denotes the number of molecules transmitted and received during the current symbol duration, and $N_{p}$ is the number of molecules transmitted from the previous symbol duration but received during the current symbol duration. For simplicity, we use $p_1$ for $P_{hit}$ during $T_s$ and $p_2$ for $P_{hit}$ during $2T_s$.

\subsection{Energy Model}
We use an energy model to that described in~\cite{Kuran_NCN10}. To prevent them from interacting with others during propagation, the messenger molecules are encapsulated in vesicles. The vesicles are then carried to the boundary of the machine (e.g. the eukaryotic cell), and released into the propagation medium. Here we use $r_{unit}$ to denote the machine radius size. 

  \begin{figure*}[!t]
 \centerline{\resizebox{1.8\columnwidth}{!}{\includegraphics{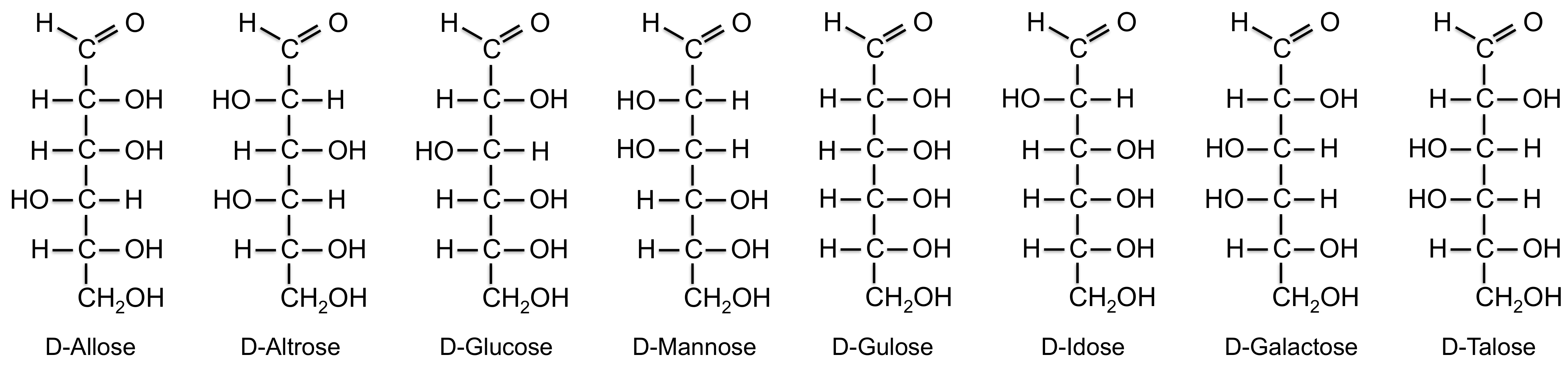}}}
  \caption{$D$-form isomers of aldohexoses.}
  \label{Fig:hexose}
\end{figure*}

The energy needed in these steps are calculated. Moreover, we can obtain the cost of synthesizing messenger molecules (hexoses as an example) by the enthalpy of formation $\Delta H$. Then the final energy model can then be presented as follows~\cite{Kuran_NCN10}:
\begin{align}
E_T& = nE_S+\frac{n}{c_v}(E_V+E_C+E_E),\nonumber\\
E_S&= \frac{\Delta H_{hexose}}{6.02\times10^{23}}  J\text{ per messenger molecule},\nonumber\\
E_V&=83\times5(4\pi{r_v}^2) zJ,\nonumber\\
E_C&=83\times \left[\frac{r_{unit}/2}{8}\right]zJ,\nonumber\\
E_E&=83\times10 zJ,\nonumber\\
c_v&={\left(\frac{r_v}{r_{mm}\sqrt{3}}\right)}^3\nonumber
\end{align}
where, $E_T$ is the total energy cost required to transmit $n$ number of molecules, $E_S$ is the synthesizing cost of one hexose molecule calculated from the sum of bond energy (e.g., the enthalpy change), $E_V$ is the vesicle-synthesizing cost having a radius of $r_v$. $E_C$ is the cost of intra-cellular transportation, $E_E$ is for membrane fusion, and $c_v$ is the capacity of one vesicle that is related to the radius of messenger molecules $r_{mm}$. 

\section{Modulation Techniques}
\label{Sec:ModTech}
In molecular communication via diffusion, the unique properties of the messenger molecules can determine modulation techniques. Two such techniques that have been proposed include the use of concentration and type of messenger molecules\cite{Kuran_ICC11}. A usable messenger molecule itself, however, has yet to be proposed. Thus this paper proposes practical messenger molecules, as well as analyzes and compares their achievable rates by applying different modulation techniques. 

\begin{figure}[!t]
 \centerline{\resizebox{0.44\columnwidth}{!}{\includegraphics{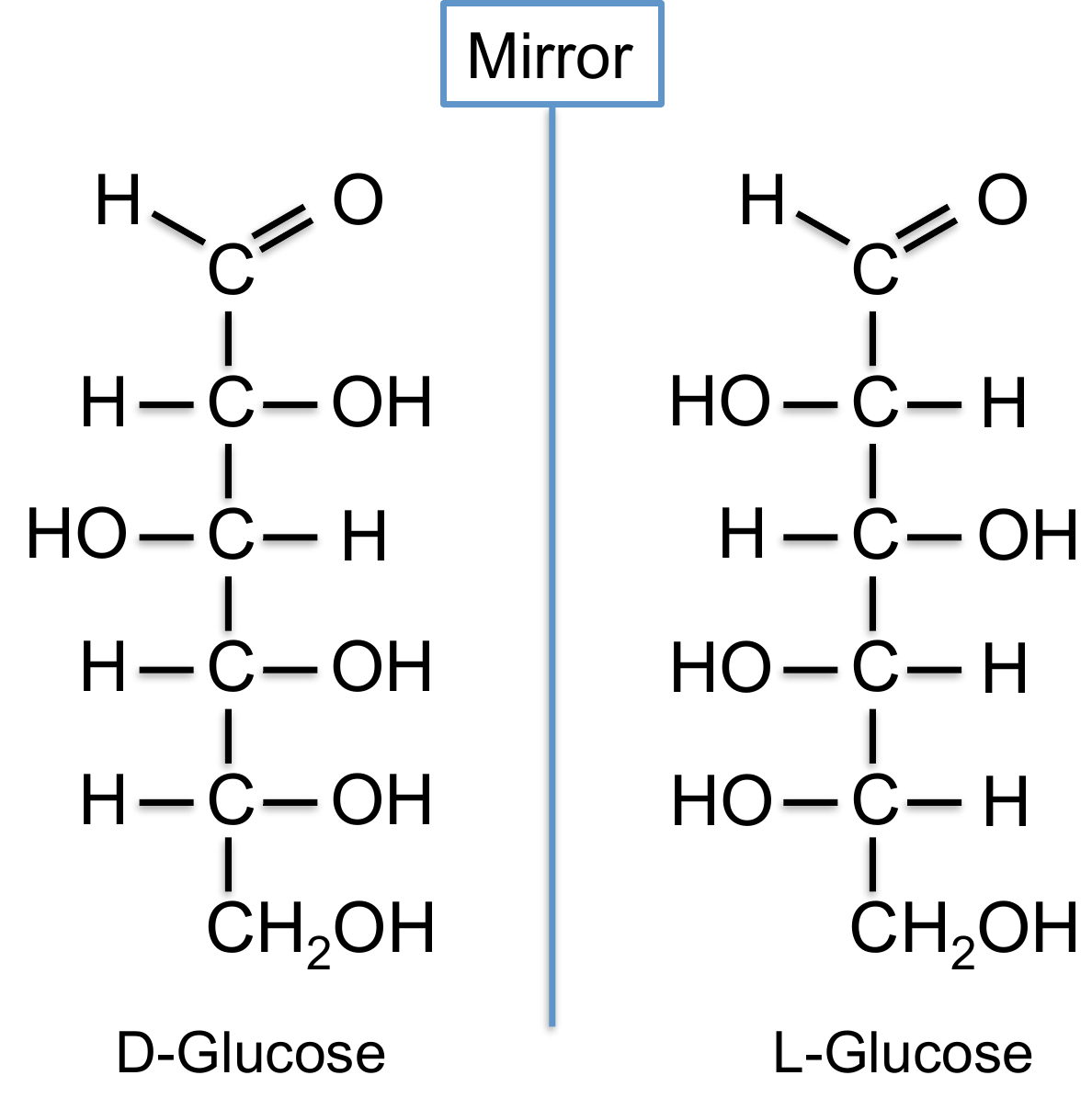}}}
  \caption{Mirror image of $D$- and $L$- glucose. They are enantiomers.}
  \label{Fig:mirror}
\end{figure}

\subsection{Isomers for Messenger Molecules}  
The most important thing we have to be careful when designing messenger molecules is that has to be non-toxic to human body. The messenger molecule suggested in \cite{Kuran_NCN10}, however, is highly flammable (i.e., hydrofluorocarbon), which means it may be inappropriate as a messenger molecule. 

For several reasons, potential candidates for messenger molecules are isomers, molecules with the same number and types of atoms \cite{BIOCHEM}. First of all, they consist of the same type of atoms, lightening the burden on the transmitter nanomachine synthesizing the messengers. For numerical analysis, this paper uses the isomers known as hexoses, especially aldohexoses. Note that we can select from the aldoses family, hexoses, pentoses, tetroses, or trioses based on the required modulation order.

Aldohexoses are monosaccharides with the chemical formula $C_6H_{12}O_6$. They have four chiral carbon atoms that give them $16$ (=$2^4$) stereoisomers. Fig.~\ref{Fig:hexose} illustrates eight kinds of $D$- form diastereomers. The enantiomers of each molecule are another set of eight $L$- form diastereomers. Therefore, aldohexoses have 16 different shapes. Here, diastereomers are stereoisomers that are not enantiomers (mirror-image isomers). For example, $D$-glucose and $L$-glucose as shown in Fig.~\ref{Fig:mirror} are enantiomers. $D$-glucose and $D$-galactose, isomers but not mirror images, are diastereomers.

\begin{figure}[!t]
 \centerline{\resizebox{0.9\columnwidth}{!}{\includegraphics{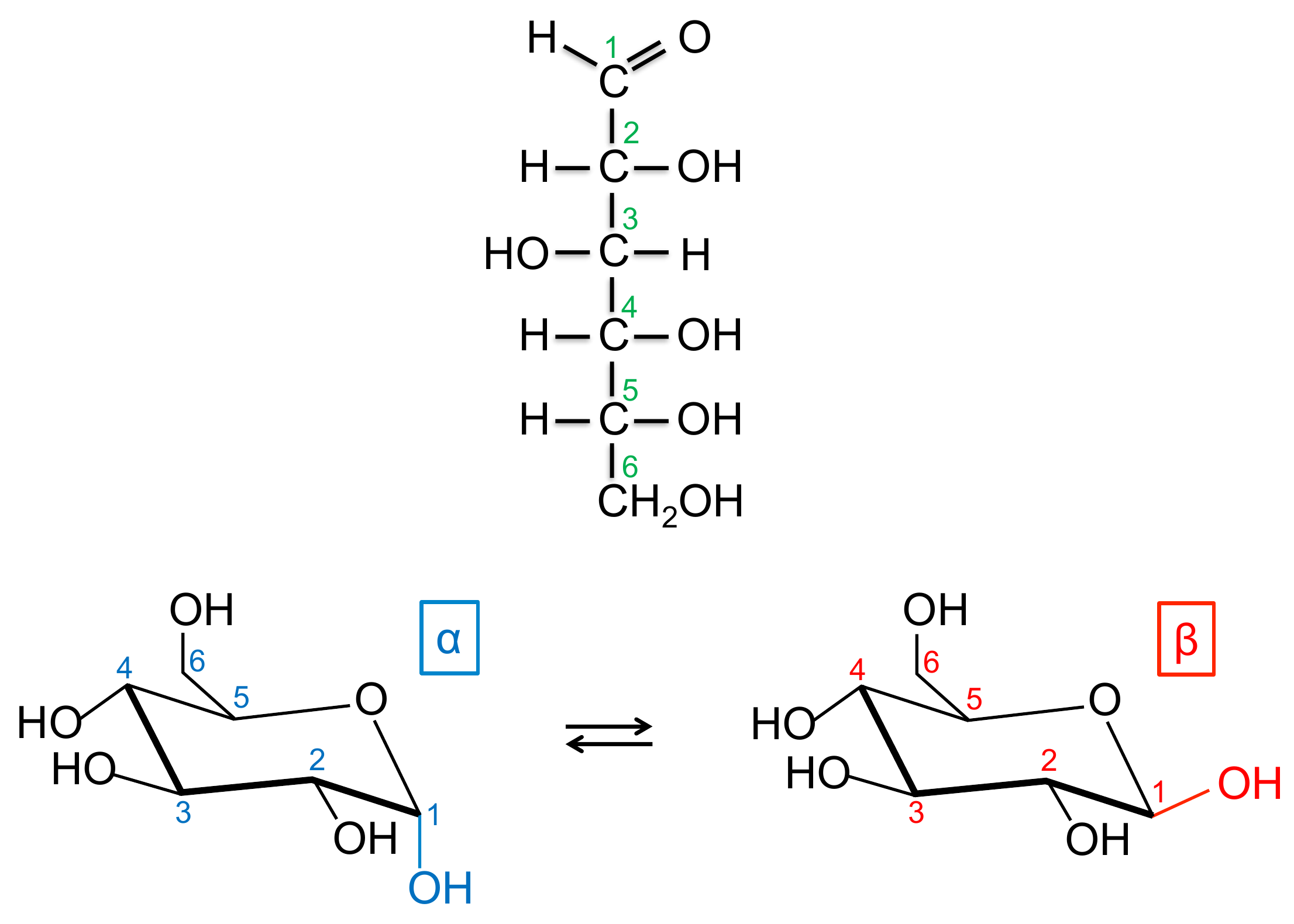}}}
  \caption{$D$-glucose and its $\alpha$- and $\beta$- anomers. They undergo rapid interconversion in water, which is called mutarotation.}
  \label{Fig:muta}
\end{figure}

When each isomer is dissolved in aqueous solution, it mostly exists as a cyclic form. $D$-glucose, for instance, undergoes nucleophilic addition reaction generating four cyclic anomers, $\alpha$-, $\beta$- forms of $D$-~glucopyranose and $D$-glucofuranose~\cite{nucleo}. We consider, however, only two pyranose forms since they predominate with $36$ and $64$ percentages. If the functional groups attached to carbon number~1 ($C1$) and the $C5$ shown in Fig.~\ref{Fig:muta} have a trans-structure, it is called $\alpha$- form, and if a cis-structure, $\beta$- form. They interconvert each other in solution through a process called mutarotation. Therefore, by deploying hexoses group in the system, there are a total of 32 different isomers. This means that we can increase a modulation order up to 32, i.e., 5 bits per symbol. 

\subsection{Molecular Concentration Based}
When the concentration of messenger molecules is used, the technique is known as concentration-based modulation, originally introduced in~\cite{Kuran_ICC11}. It is also referred to as CSK (concentration shift keying). In binary CSK (BCSK), one threshold exists, and a receiver nanomachine decodes the symbol as `1' if the number of received messenger molecules exceeds the threshold, `0' otherwise~\cite{Kuran_NCN10}. Generally, 2$^n$-CSK systems transmit $n$ bits per symbol, and requires 2$^n$-1 number of thresholds. In this system, there is, theoretically, no limit in the modulation order. As the modulation order increases, however, so does the error probability since the minimum distance between two neighboring thresholds decreases. It only uses one kind of molecule, and $D$-glucopyranose is used for analysis here. Below is a probabilistic analysis for BCSK systems:
\begin{align}\begin{split}&
P_a(0,0) = P_b(0,0,0) + P_b(1,0,0) \\
&= \frac{1}{4}\left[P(N_n<\tau)+P(N_p+N_n<\tau)\right] \\
&= \frac{1}{4}\Bigg[2-Q\left(\frac{\tau}{\sigma}\right)-Q\left(\frac{\tau-n(p_2-p_1)}{\sqrt{n[p_2(1-p_2)+p_1(1-p_1)]+\sigma^2}}\right) \Bigg],\end{split}\nonumber\end{align}
\begin{align}\begin{split}
&P_a(0,1) = P_b(0,0,1) + P_b(1,0,1) \\
&= \frac{1}{4}\left[P(N_n\geq\tau)+P(N_p+N_n\geq\tau)\right] \\
&= \frac{1}{4}\Bigg[Q\left(\frac{\tau}{\sigma}\right)+Q\left(\frac{\tau-n(p_2-p_1)}{\sqrt{n[p_2(1-p_2)+p_1(1-p_1)]+\sigma^2}}\right) \Bigg]  \nonumber\end{split}
\end{align}
where, $P_a(X,Y)$ indicates the probability of $X$ sent and $Y$ received, and $P_b(Z,X,Y)$ is the probability of $X$ sent, $Y$ received, and $Z$ previously sent. $Q(\cdot)$ is the tail probability of the normal distribution.

\subsection{Molecular Type Based}
When different types of molecules represent different symbols, the technique is known as molecular-type-based modulation, referred to as molecule shift keying (MoSK). Unlike the work in \cite{Kuran_ICC11}, we here use a set of isomers. We thus name it IMoSK (isomer-based MoSK). IMoSK requires only one threshold, making it simpler than the CSK system. For systematic analysis, we can choose one among the several aforementioned isomer sets; a modulation order can be determined by the sets used. For example, hexoses have a modulation order of up to $32$, trioses of $4$. For simplicity, we apply, as did in~\cite{Kuran_ICC11}, the addictive white Gaussian noise (AWGN) model. The mutarotation effect is also considered. 
\subsubsection{B-IMoSK-AWGN}
Only AWGN considered,
\begin{align}\begin{split}
P_a(\alpha,\alpha)_{AWGN} &= P_b(\alpha,\alpha,\alpha) + P_b(\beta,\alpha,\alpha) \\
&=\frac{1}{4}[P(N_p+N_c+N_n\geq\tau)P(N_n<\tau)\\
&~+P(N_c+N_n\geq\tau)P(N_p+N_n<\tau)],\nonumber\end{split} \\ \begin{split}
P_a(\alpha,\beta)_{AWGN} &= P_b(\alpha,\alpha,\beta) + P_b(\beta,\alpha,\beta) \\
&=\frac{1}{4}[P(N_n\geq\tau)P(N_p+N_c+N_n<\tau)\\
&~+P(N_p+N_n\geq\tau)P(N_c+N_n<\tau)].\nonumber\end{split} \end{align}
Note that $P_a(\beta, \beta)$ and $P_a(\beta, \alpha)$ can also be calculated similarly. Due to the page limit, we omit it here.

\subsubsection{B-IMoSK-Mutarotation Considered ($\alpha$-$D$-glucopyranose $\Leftrightarrow$ $\beta$-$D$-glucopyranose)}
When $\alpha$- and $\beta$- $D$-glucopyranose are chosen, there is a possibility of incorrect decoding, such as $\alpha$- sent, $\beta$- received, or  $\beta$- sent, $\alpha$- received due to mutarotation. Thus we derive an error probability considering the mutarotation process. The $\alpha$- or $\beta$- form sent varies its number with time\cite{muta}, and the number can be calculated by observing the specific optical rotation. $R_{t\alpha}$ or $R_{t\beta}$ (observed specific optical rotation at time $t$) minus $R_{eq}$ (at equilibrium) divided by $R_{\alpha}$ or $R_{\beta}$ (at time $0$) minus $R_{eq}$ has a linear relationship with time ($T_s$) at about 36.5$^{\circ}C$. $R_{t\alpha}$ and $R_{t\beta}$ values are calculated by $R_{\alpha}$ and $R_{\beta}$, and $R_{eq}$ cab be found in \cite{muta}. The number of $\alpha$- and $\beta$ form existed after time $T_s$ is obtained as below. If the number of $\beta$- form exceeds the threshold value after $T_s$ when $\alpha$- sent, $\frac{\beta}{n}$ value is added to the error term. 
Thus, $P_a (\alpha, \beta)$ can be calculated as
\begin{align}
&\frac{R_{t\alpha}-R_{eq}}{R_\alpha-R_{eq}}=-\frac{0.99}{3600}t_s+1,   \nonumber\\ 
&R_{t\alpha}=\frac{{\alpha}R_{\alpha}+{\beta}R_{\beta}}{\alpha+\beta}=\frac{{\alpha}R_{\alpha}+(n-\alpha)R_{\beta}}{n},  \nonumber \\ 
&\alpha=\frac{(R_{t\alpha}-R_{\beta})n}{R_{\alpha}-R_{\beta}}, ~~\beta=n-\alpha.   \nonumber\\
&\text{If } \beta\geq\tau,~P_a(\alpha,\beta)=P_a(\alpha,\beta)_{AWGN}+\frac{\beta}{n} \nonumber
\end{align}

where, $R_{eq}$ = 52.7$^\circ$, $R_{\alpha}$=112.2$^\circ$, and $R_{\beta}$=18.7$^\circ$~\cite{muta}. The parameters $\alpha$ and $\beta$ indicate the number of $\alpha$- form and $\beta$- form molecules, respectively, and $n$ is the number of total molecules transmitted. Optical specific rotation of the chemical compound is defined as the observed angle of optical rotation when plane-polarized light passes through $D$-glucopyranoses.\footnote{It could be measured by a polarimeter, and there is a linear relationship between the observed rotation and the concentration of the compound~\cite{OCHEM}}
\begin{table}[!t]
\caption{Simulation Parameters.}
\begin{center}
\begin{tabular}{|c|c|}
\hline
Parameters & Values \\
\hline\hline
$P_{hit}$ for $T_s$ & 0.6097 \\ \hline
$P_{hit}$ for $2T_s$ & 0.7208 \\ \hline
Radius of the hexoses~\cite{hexose_radi} & 0.38 $nm$ \\ \hline
$D$ of hexoses (\ref{D}) &597.25 ${\mu m}^2/sec$ \\ \hline
$\triangle H_{\text{hexose}}$ & 1271 $kJ/mol$\\ \hline
$T_s$ \cite{Kuran_NCN10}& 5.9$sec$ \\ \hline
Viscosity of the water & 0.001$kg/sec \cdot m$\\ \hline
Temperature (body temperature) & 36.5 $^{\circ}C$ = 310$K$ \\ \hline
\end{tabular}
\end{center}
\label{parameters}
\end{table}

\subsubsection{32-IMoSK}
When using hexoses, the system has the maximum modulation order of $32$. Probabilities of $X$ sent and $Y$ received for all $X$ and $Y$ values are obtained similarly and thus omitted here (see~\cite{CNT_Kim12}).

\section{Numerical Results}
\label{Sec:Num}
Assume that all hexoses have the same physical properties; size, diffusion coefficient, and enthalpy of formation.
$P_{hit}$ value is calculated by the same numerical calculation used in \cite{Kuran_NCN10}. We approximate the value for the hexoses by assuming it to be proportional to the diffusion coefficient. Therefore, we have the hitting probabilities shown in Table~\ref{parameters} and apply these into the proposed scheme explained in Section~\ref{Sec:ModTech}.
Here, we define the achievable rate $R$ that maximizes the mutual information $I(X;Y)$ as follows
\begin{align}\begin{split}
&I(X;Y)=\sum P(X,Y)\log_2{\frac{P(X,Y)}{P(X)P(Y)}}, \\
&R =\max_{\tau} I(X;Y) \end{split}  \label{cap}
\end{align}
where, $P(X, Y)$ denotes the joint probability of $X$ and $Y$ while $P(X)$ and $P(Y)$ are the probabilities of events $X$ and $Y$. In (\ref{cap}), the threshold values vary from 1 to 1000.

\begin{figure}[!t]
 \centerline{\resizebox{0.85\columnwidth}{!}{\includegraphics{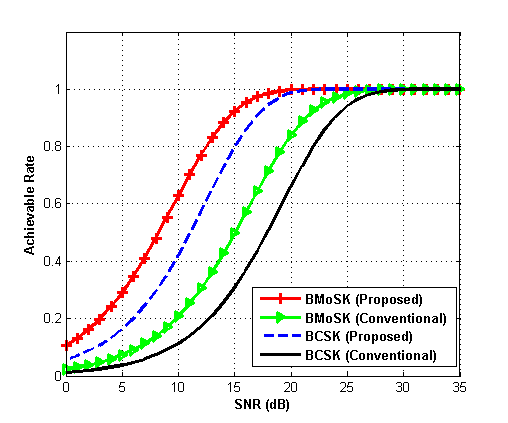}}}
  \caption{Achievable rate comparisons of the conventional insulin based and the proposed methods using two kinds of aldohexose isomers.}
  \label{Fig:B_comparison}
\end{figure}
\begin{figure}[t]
 \centerline{\resizebox{0.85\columnwidth}{!}{\includegraphics{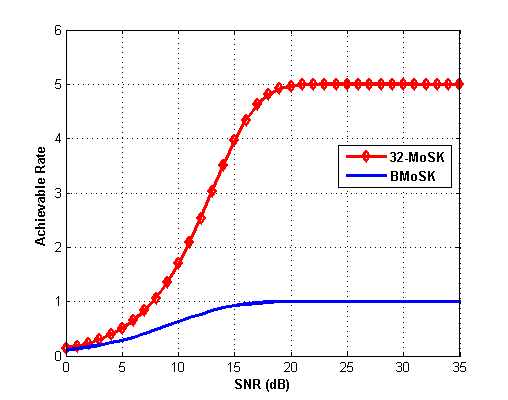}}}
  \caption{Achievable rates of 32-IMoSK and B-IMoSK. Thirty two kinds of isomers of aldohexose used for 32-IMoSK.}
  \label{Fig:MoSK32}
\end{figure}

Fig.~\ref{Fig:B_comparison} compares the achievable rate of the proposed method using hexoses with the conventional insulin based method in BCSK and B-IMoSK systems. It is remarkable that we obtain about 8~dB of SNR gain in both BCSK and B-IMoSK systems. In addition, the B-IMoSK system shows a better data rate performance than BCSK. This is due mainly to the size of the proposed messenger molecule being much smaller than that of insulin. Hence, the transmit energy of the proposed method is much less than that of the conventional method. This result will be discussed in more detail in~\cite{CNT_Kim12}.

\begin{figure}[t]
 \centerline{\resizebox{0.85\columnwidth}{!}{\includegraphics{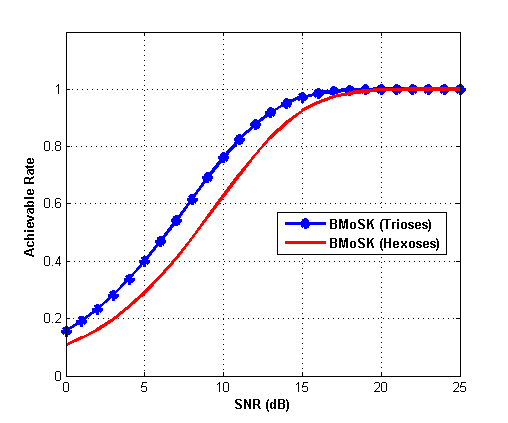}}}
  \caption{Achievable rate comparisons. Triose and hexose used.}
  \label{Fig:BMoSK_tri_hex}
\end{figure}

Fig.~\ref{Fig:MoSK32} shows the achievable rates of 32-IMoSK and B-IMoSK using hexoses as messenger molecules. As can be seen from Fig.~\ref{Fig:MoSK32}, the 32-IMoSK system has the maximum rate of 5 (bits per symbol), and the B-IMoSK of 1 (bit per symbol).  In Fig.~\ref{Fig:BMoSK_tri_hex}, we compare the achievable rates using trioses and hexoses. Obviously, trioses achieve higher SNR gain than hexoses due to their smaller sizes. Trioses, however, have the data transmission limit  1 bit per symbol. From this result, we can conclude that trioses can be selected for a low data rate system with a higher transmission reliability and hexoses for a high data rate system. 


\section{Conclusion}
\label{Sec:Conc}
To make nano communication feasible in practice, this work proposed novel modulation techniques using isomers as messenger molecules. We first introduced energy and channel models for our system. Next we proposed several modulation methods able to support up to 5 bits per symbol. We also compared the achievable rate performance with existing modulation methods (concentration-based and molecular-type-based). This work differs from prior work in that it proposes practical messenger molecules and provides guidelines for selecting from among several possible candidates. Future work, we will consider, to achieve more modulation degrees of freedom, the ratio of enatiomers. To increase the transmission data rate further, we will also consider multiple sets of messenger molecules.

\bibliographystyle{IEEEtran}

\bibliography{references_monacom}

\end{document}